\documentclass[a4paper,fleqn,usenatbib]{mnras} 
\usepackage{newtxtext,newtxmath}
\usepackage[T1]{fontenc}
\usepackage{ae,aecompl}

\newcommand{\beq}{\begin{equation}}
\newcommand{\eeq}{\end{equation}}
\newcommand{\bal}{\begin{aligned}}
\newcommand{\eal}{\end{aligned}}
\newcommand{\beqa}{\begin{equation}\begin{aligned}}
\newcommand{\eeqa}{\end{aligned}\end{equation}}
\newcommand{\avg}[1]{\left\langle{#1}\right\rangle}

\newcommand{\Msun}{\rm M_\odot}
\newcommand{\Lsun}{\rm L_\odot}

\newcommand{\fcen}{{f_\nu^{\rm cen}}}
\newcommand{\fsat}{{f_\nu^{\rm sat}}}
\newcommand{\fcenp}{{f_{\nu'}^{\rm cen}}}
\newcommand{\fsatp}{{f_{\nu'}^{\rm sat}}}
\newcommand{\jnu}{\bar{j}_{\nu}}
\newcommand{\jnup}{\bar{j}_{\nu'}}

\newcommand{\fsky}{f_{\rm sky}}
\newcommand{\FoM}{\rm FoM}

\newcommand{\sigmapix}{\sigma_{\rm pix}}
\newcommand{\thetaFWHM}{\theta_{\rm FWHM}}
\newcommand{\Nfreq}{N_{\rm freq}}

\newcommand{\MJysr}{{\rm MJy\ sr^{-1}}}
\newcommand{\uKarcmin}{{\rm \mu K \mbox{-} arcmin}}
\newcommand{\uKCMBarcmin}{{\rm \mu K_{CMB} \mbox{-} arcmin}}
\newcommand{\uKCMB}{{\rm\mu K_{CMB}}}

\newcommand{\LIR}{L_{\rm IR}}
\newcommand{\SFR}{{\rm SFR}}

\newcommand{\Lmax}{\ell_{\rm max}}
\newcommand{\numax}{\nu_{\rm max}}
\newcommand{\numin}{\nu_{\rm min}}

\newcommand{\Mmin}{{M_{\rm min}}}

\newcommand{\dM}{{\rm d}M}
\newcommand{\dN}{{\rm d}N}
\newcommand{\dn}{{\rm d}n}
\newcommand{\dS}{{\rm d}S}
\newcommand{\dV}{{\rm d}V}
\newcommand{\dz}{{\rm d}z}
\newcommand{\rmd}{{\rm d}}

\usepackage{mathtools}
\usepackage{multirow, booktabs} 
\usepackage{amsmath,amssymb}

\title[Optimizing Future CFIRB Experiments]{
Optimizing future experiments of cosmic far-infrared background: a principal component approach}
\author[Wu and Dor\'e]{Hao-Yi Wu$^{1,2}$\thanks{E-mail: hywu@caltech.edu} and Olivier Dor\'e$^{1,2}$\\
$^{1}$California Institute of Technology, MC 367-17, Pasadena, CA 91125, USA \\
$^{2}$Jet Propulsion Laboratory, California Institute of Technology, 4800 Oak Grove Drive, Pasadena, CA 91109, USA}
\date{Accepted 2017 February 9. Received 2016 December 8; in original form 2016 December 8}
\pubyear{2017}
\begin{document}
\label{firstpage}
\pagerange{\pageref{firstpage}--\pageref{lastpage}}
\maketitle

\begin{abstract}
The anisotropies of cosmic far-infrared background (CFIRB) probe the star formation rate (SFR) of dusty star-forming galaxies as a function of dark matter halo mass and redshift.  We explore how future CFIRB experiments can optimally improve the SFR constraints beyond the current measurements of {\em Planck}.  We introduce a model-independent, piecewise parametrization for SFR as a function of halo mass and redshift,  and we calculate the Fisher matrix and principal components of these parameters to estimate the SFR constraints of future experiments.  We investigate how the SFR constraints depend on angular resolution, number and range of frequency bands, survey coverage, and instrumental sensitivity.  We find that the angular resolution and the instrumental sensitivity play the key roles.  Improving the angular resolution from 20 to 4 arcmin can improve the SFR constraints by 1.5--2.5 orders of magnitude. With the angular resolution of {\em Planck}, improving the sensitivity by 10 or 100 times can improve the SFR constraints by one or two orders of magnitude, and doubling the number of frequency bands can also improve the SFR constraints by an order of magnitude.  We find that survey designs like the {\em Cosmic Origins Explorer} ({\em CORE}) are very close to the optimal design for improving the SFR constraints at all redshifts, while survey designs like {\em LiteBIRD} and CMB-S4 can significantly improve the SFR constraints at $z\gtrsim3$.

\end{abstract}


\section{Introduction}

Cosmic far-infrared background (CFIRB)\footnote{We use the longer acronym CFIRB instead of CIB, because the latter sometimes refers to cosmic near-infrared background, which originates from old stellar populations.} originates from unresolved, dusty star-forming galaxies across cosmic time. In dusty star-forming galaxies, newly-formed massive stars produce abundant ultraviolet (UV) photons, and $\sim$ 90\% of these UV photons are absorbed by the interstellar dust, which is heated to approximately 15--60 K and emits nearly blackbody radiation in infrared (IR), with a peak at $\sim$ 60--100 $\micron$ (3000--5000 GHz) in the rest frame.  At high redshift, these galaxies are observed in far-infrared (FIR) and submillimetre (submm) bands.  These FIR/submm observations of galaxies probe their star formation rate (SFR) and are highly complementary to UV observations \citep[e.g.,][]{Casey14,Lutz14,MadauDickinson14}. However, the majority of the FIR/submm galaxies are unresolved due to the low resolution of telescopes at these wavelengths. Therefore, CFIRB provides a unique way to reveal the star-forming activities under the current resolution limit. In particular, the anisotropies of CFIRB probe how SFR is related to the underlying dark matter haloes.

The observability of CFIRB was first predicted by \cite{Bond86}.  Since its discovery by {\em COBE} \citep{Puget96,Fixsen98,Hauser98,HauserDwek01}, and subsequent observations of {\em ISO} \citep{Elbaz02} and {\em Spitzer} \citep{Dole06,Lagache07}, CFIRB has opened a new window for observing the star formation activities in high-redshift, low-mass galaxies \citep[e.g.,][]{CharyElbaz01,Lagache05}.  More recently, the measurements of CFIRB anisotropies have been significantly improved by
BLAST \citep{Viero09},
SPT \citep{Hall10}, 
{\em AKARI} \citep{Matsuura11}, 
ACT \citep{Hajian12}, 
{\em Herschel} \citep{Amblard11,Berta11,Viero13},
and {\em Planck} \citep{Planck11cib, Planck13XXX}.
These measurements have enabled detailed modelling of the galaxy populations contributing to CFIRB \citep[e.g.,][]{DeBernardis12,Shang12,Xia12, Addison13,Bethermin13,Thacker13,Planck13XXX,Cowley16, Wu16b,Wu16}.

The next-generation space-borne missions of cosmic microwave background (CMB) will include FIR/submm bands to measure dust emissions
and will significantly improve CFIRB measurements. These missions include 
the {\em Cosmic Origins Explorer} ({\em CORE}; \citealt{CORE16a,CORE16b}), 
the {\em Primordial Inflation Explorer} ({\em PIXIE}; \citealt{Kogut11}), 
the {\em LiteBIRD} \citep{LiteBIRD} and
the Experimental Probe of Inflationary Cosmology (EPIC, \citealt{EPIC}).
In addition, the next-generation ground-based CMB-S4 experiment \citep{Abazajian16} will have superior resolution and sensitivity, which are essential for improving CFIRB measurements. The CFIRB measurements from these missions will have the potential to constrain the cosmic star-formation history to high accuracy.  With the planning of these missions underway, it is imperative to understand the optimal survey designs for constraining the cosmic star-formation history. 

In this work, we explore how future CFIRB experiments can most effectively constrain the cosmic star-formation history.  We introduce a model-independent, piecewise parametrization for SFR of galaxies as a function of halo mass and redshift, and we calculate the Fisher matrix and principal components of these parameters to assess the SFR constraints. We explore the impact of angular resolution, number and range of frequency bands, survey coverage, and instrumental sensitivity. We find that the angular resolution and the instrumental sensitivity are the two main factors for determining the SFR constraints from an experiment. For example, an angular resolution improvement from 20 to 4 arcmin can lead to 1.5--2.5 orders of magnitude improvement in SFR constraints. With the angular resolution of {\em Planck}, increasing the sensitivity by 10 times (as in the case of {\em CORE}) or 100 times (as in the cases of {\em LiteBIRD} and CMB-S4) with respect to {\em Planck} will lead to one or two orders of magnitude improvement in SFR constraints. 

In addition, we find that increasing the number of frequency bands is not always effective in improving SFR constraints. For example, with the angular resolution of {\em Planck}, doubling the number of frequency bands of {\em Planck} can improve the SFR constraints by approximately an order of magnitude, but further increasing the number of bands has less significant impact.  However, if we degrade the angular resolution to 20 arcmin, increasing the number of bands leads to limited improvement in SFR constraints. Comparing our results with the designs of several future surveys, we find that the survey designs of {\em CORE} are very close to optimal for constraining SFR at all redshifts, while the survey designs of {\em LiteBIRD} and CMB-S4 can provide valuable SFR constraints at $z\gtrsim3$.

This paper is organized as follows. 
In Section \ref{sec:model}, we review the halo model for calculating the CFIRB angular power spectra. 
In Section \ref{sec:pca}, we introduce our model-independent, piecewise parametrization for SFR,
as well as the Fisher matrix and the principal component approach.
Section~\ref{sec:fom} explores how SFR constraints depend on angular resolution, number of frequency bands, sky coverage, and instrumental sensitivity.
Section~\ref{sec:comb} studies the optimal range of frequency bands.
We discuss the implications of our results for future missions in Section~\ref{sec:discussions}, and 
we summarize in Section \ref{sec:summary}.

Following \cite{Planck13XXX}, we adopt a flat $\Lambda$CDM cosmology with the following cosmological parameters:
$\Omega_{\rm m} = 0.3175$
$\Omega_{\Lambda} = 0.6825$;
$\Omega_{\rm b} h^2 = 0.022068$;
$\sigma_8 = 0.8344$.
$h = 0.6711$; 
$n_{\rm s} = 0.9624$.
The halo mass in this work refers to the virial mass.

\section{Modelling the CFIRB Power Spectra}\label{sec:model}

\begin{figure}
\centering
\includegraphics[width=1\columnwidth]{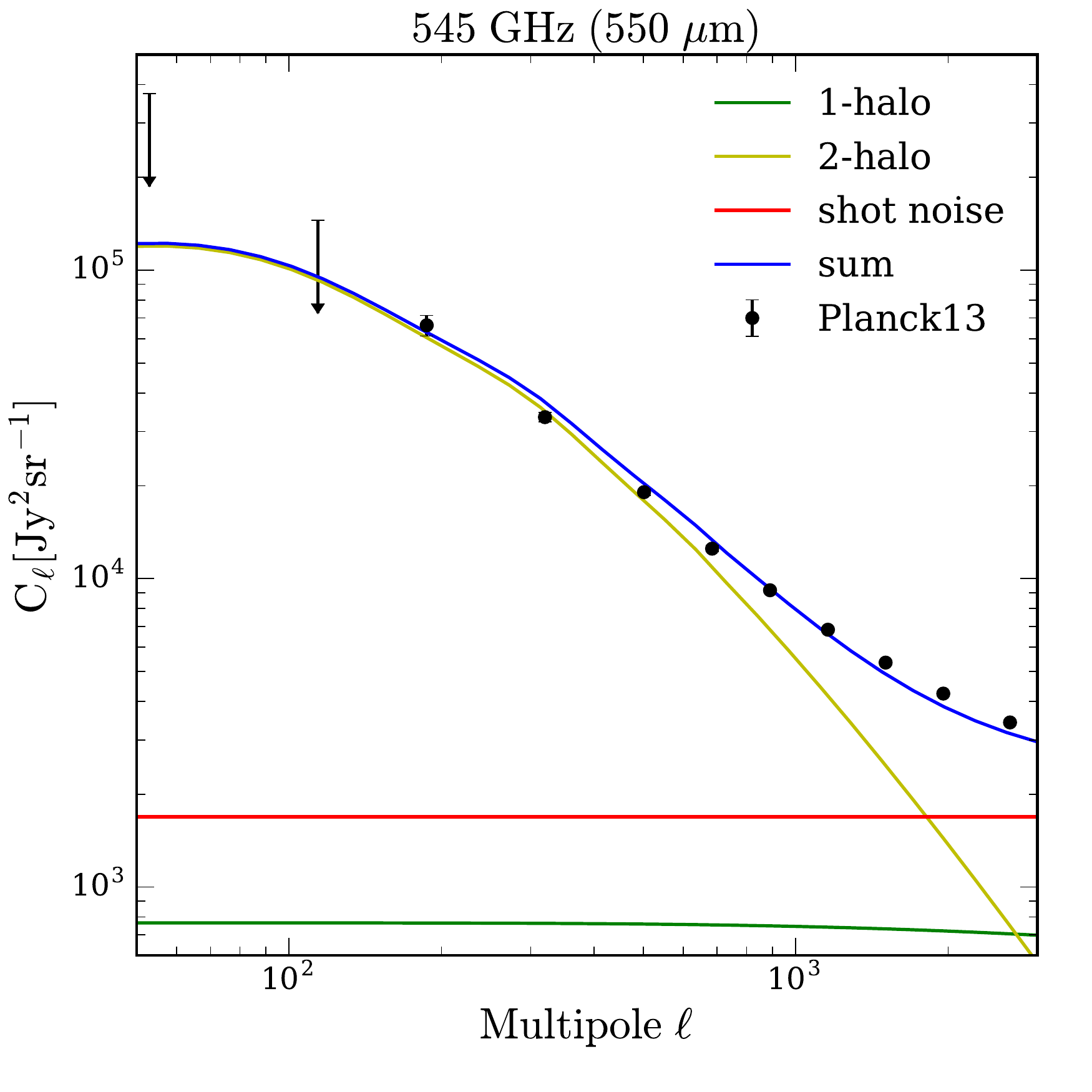}
\caption{Angular power spectrum of CFIRB at 545 GHz (550 $\micron$).  The theory prediction (blue) is broken down into the 1-halo term (green), the 2-halo term (yellow), and the shot noise (red).  The black points are the observational results from P13 (given by their table D.2).}
\label{fig:CL_demo}
\end{figure}

We adopt the halo model implementation in \citet[][S12 thereafter]{Shang12} and in the {\em Planck} 2013 results \citep[][P13 thereafter]{Planck13XXX}  as our fiducial model to calculate the angular power spectra of CFIRB.  This particular choice does not impact our results because we assess the SFR constraints in a model-independent way (see Section~\ref{sec:pca}).\footnote{In \cite{Wu16} and \cite{Wu16b}, we develop physical and empirical models for interpreting CFIRB and other FIR/submm observables. These models help us interpret CFIRB physically and explore the consistency between FIR, optical, and UV observations.}  Our calculation includes the following steps:
\begin{itemize}

\item modelling the SFR and the IR spectral luminosity as a function of halo mass and redshift (Section~\ref{sec:Lnu}),

\item calculating the 1-halo and 2-halo contributions to the CFIRB power spectra (Section~\ref{sec:CL}),

\item calculating the shot noise (Section~\ref{sec:shot}),

\item estimating the detector noise (Section~\ref{sec:detector}).

\end{itemize}
Figure~\ref{fig:CL_demo} demonstrates the halo model prediction of the CFIRB angular power spectrum at 545 GHz (550 $\micron$), which agrees well with the observational results in P13. Below we describe our implementation in detail.

\subsection{IR luminosities and spectral energy distributions of galaxies}
\label{sec:Lnu}

One of the main goals of CFIRB experiments is to constrain SFR as a function of halo mass and redshift.  We adopt the common assumption that the SFR is proportional to the IR luminosity,
\beq
\SFR(M,z) = K \LIR(M,z) \ , 
\eeq
where $K = 1.7 \times10^{10} \Msun \rm yr^{-1} \Lsun$ based on the Salpeter initial mass function \citep{Kennicutt98}.  Following P13, we assume that $\LIR(M,z)$ is parametrized as
\beq
\LIR(M,z) = L_0 \Phi(z) \Sigma(M) \ ,
\eeq
where the redshift dependence is given by
\beq
\Phi(z) = (1+z)^\delta \ ,
\eeq
and the mass dependence is given by 
\beq
\Sigma(M) = \frac{M}{(2{\rm\pi}\sigma_{M}^2)^{1/2}} \exp\left\{ - \frac{(\log_{10}M - \log_{10}M_{\rm eff})^2}{2\sigma_{M}^2} \right\} .
\label{eq:SigmaM}
\eeq
We adopt the best-fitting values in table 9 in P13: 
$\delta = 3.6$;
$M_{\rm eff} = 10^{12.6}\ \Msun$;
$\sigma^2_M = 0.5$; 
$L_0=0.0135\ \Lsun$.

To predict the observed CFIRB, we need the spectral luminosity of galaxies, $L_{\nu}(M,z)$.  We assume that the spectral energy density (SED), $\Theta_\nu$, depends only on redshift and is independent of $M$ and $\LIR$; thus, 
\beq
L_{\nu}(M,z) = \LIR(M,z) \Theta_\nu(z) \ . 
\eeq
We assume that the SED is parametrized separately for low frequency (modifying the Rayleigh--Jeans law) and high frequency (modifying the Wien's law to a shallower power-law), 
\beq\label{eq:SED}
\Theta_\nu \propto \left\{
\begin{array}{ll} 
\nu^\beta B_\nu(T_{\rm d}) & \mbox{for $\nu < \nu_0$}\\
A \nu^{-\gamma} & \mbox{for $\nu \geq \nu_0$}
\end{array} \right. \ , 
\eeq
where $B_\nu$ is the Planck function.   The factor $A$ and the peak frequency $\nu_0$  are calculated by imposing the continuity of $\Theta(\nu)$ and ${\rm d}\Theta/{\rm d}\nu$ at $\nu=\nu_0$.  The SED is normalized such that $\int \Theta_\nu d\nu = 1$; we integrate between 100 and $10^6$ GHz, which has been tested to ensure convergence.  We assume that the dust temperature $T_{\rm d}$  depends only on redshift, 
\beq
T_{\rm d} = T_0 (1+z)^\alpha \ .
\eeq
We again adopt the best-fitting values in table 9 in P13:
$T_0 = 24.4$ K;
$\alpha = 0.36$;
$\beta = 1.75$;
$\gamma = 1.7$.

\subsection{CFIRB angular power spectra}\label{sec:CL}

The mean emission coefficient of CFIRB is given by
\beq
\jnu(z) =\int \dM \frac{\dn}{\dM} \big[ \fcen(M,z) + \fsat(M,z) \big] \ , 
\eeq
where $\fcen$ and $\fsat$ correspond to the contribution from central and satellite galaxies, respectively, 
\beqa
\fcen(M,z) &= \frac{1}{4{\rm\pi}} N_cL_{(1+z)\nu}(M,z) \ , \\
\fsat(M,z) &= \frac{1}{4{\rm\pi}} \int_{\Mmin}^{M} \dM_s \frac{\dN}{\dM_s}(M) L_{(1+z)\nu}(M_s,z) \ .
\eeqa
In the equations above, $(1+z)\nu$ is the rest-frame frequency;
we assume that central and satellite galaxies follow the same $L_{\nu}(M,z)$; $\dn/{\rm d}M$ is the halo mass function, and we adopt the fitting function in \cite{Tinker10}; $N_c(M)$ determines whether a halo of mass $M$ hosts a central galaxy, and $N_c = 1$ if $M \ge \Mmin$ and 0 otherwise; we assume $\Mmin= 10^{10}\Msun$;  ${\rm d}N/{\rm d}M_s(M,z)$ is the number of subhaloes of mass $M_s$ in a central halo of mass $M$, and we adopt the fitting function in \cite{TinkerWetzel10}.

The 3-D auto and cross power spectrum of ($j_{\nu},j_{\nu'}$) is given by the sum of the 2-halo and the 1-halo terms, 
\beq
P_{\nu\nu'}(k,z) = P_{\nu\nu'}^{\rm 2h}(k,z) + P_{\nu\nu'}^{\rm 1h}(k,z) \ .
\eeq

The 2-halo term is contributed by galaxy pairs from two distinct haloes and is proportional to the linear matter power spectrum $P_{\rm lin}(k)$, 
\beq
P_{\nu\nu'}^{\rm 2h}(k,z) = \bar{b}_{\nu}(z) {\bar b}_{\nu'}(z) P_{\rm lin}(k,z) \ ,
\label{eq:P2h}
\eeq
where ${\bar b}_{\nu'}(z)$ is the effective galaxy bias calculated by integrating the halo bias $b(M,z)$,
\beq
\bar{b}_\nu = \frac{1}{\jnu}\int \dM \frac{\dn}{\dM} b(M,z) \big( \fcen + \fsat \big) \ .
\eeq
We use the linear matter power spectrum calculated by CAMB \citep{Lewis00} and the fitting function of halo bias in \cite{Tinker10}.

The 1-halo term is contributed by galaxy pairs in the same halo,
\beqa
P_{\nu\nu'}^{\rm 1h}(k,z) = & \frac{1}{\jnu \jnup}\int \dM \frac{\dn}{\dM} \\
& \times \big( \fcen \fsatp u + \fcenp \fsat u + \fsat \fsatp u^2 \big) \ ,
\label{eq:P1h}
\eeqa
where $u(M,k)$ is the halo mass density profile in Fourier space; we use the NFW profile \citep{Navarro97} with the concentration--mass relation from \cite{Bhattacharya13}.

With Limber approximation, the angular power spectrum of the CFIRB emission is given by
\beq
C_{\ell,\nu\nu'}^{\rm halo} = \int \frac{\dz}{\chi^2} \frac{{\rm d}\chi}{\dz} a^2 \jnu(z) \jnup(z) P_{\nu\nu'}\left(k=\frac{\ell}{\chi}, z\right) \ ,
\eeq
where $\chi(z)$ is the comoving distance. 

Figure~\ref{fig:CL_demo} shows the 1-halo and the 2-halo contributions to the angular power spectrum at 545 GHz (550 $\micron$). As can be seen, the 2-halo term dominates most of the angular scales, and the 1-halo term is lower than the shot noise at all scales.   For the CFIRB observed by {\em Planck}, the 1-halo term is subdominant for all frequency bands and angular scales (see fig.\ 12 in P13).

\subsection{Shot noise}\label{sec:shot}

The shot noise of the power spectrum corresponds to self-pairs of galaxies.  Since a galaxy can appear in different bands, the shot noise exists in both auto and cross power spectra.  The shot noise is calculated by integrating the spectral flux function (also known as the number counts), $dn/\dS_\nu$.   The spectral flux is related to the spectral luminosity via
\beq
S_\nu = \frac{L_{(1+z)\nu}}{4{\rm\pi} \chi^2 (1+z)} \ .
\eeq
The shot noise is given by
\beq
C^{\rm shot}_{\nu\nu'} = \int \dV \int \dS_\nu \frac{\dn}{\dS_\nu} S_\nu S_{\nu'} \ .
\eeq
We present the detailed derivation in Appendix~\ref{app:shotnoise}.  The red horizontal line in Figure~\ref{fig:CL_demo} demonstrates the shot noise level at 545 GHz (550 $\micron$).

\subsection{Detector noise}\label{sec:detector}

\begin{table*}
\centering
\setlength{\tabcolsep}{0.5em}
\begin{tabular}{ccccccc}
\toprule
\rule[-2mm]{0mm}{6mm} \multirow{2}[3]{*}{Frequency}& \multicolumn{4}{c}{{\em Planck}} & $0.015\ \MJysr$ and $5$ arcmin & $10^{-4}\ \MJysr$ and $5$ arcmin \\ 
\cmidrule(lr){2-5} \cmidrule(lr){6-6} \cmidrule(lr){7-7}
 & $\rm\sigma_{pix}$ & $ \rm\sigma_{pix}$ & $\thetaFWHM$ & Sensitivity & Sensitivity & Sensitivity \\
 $\rm(GHz)$ & $(\MJysr)$& $\rm(\mu K_{CMB})$& (arcmin) & ($\uKCMBarcmin$) & ($\uKCMBarcmin$) & ($\uKCMBarcmin$)\\\midrule
\rule[-2mm]{0mm}{6mm} 217 & 0.005 & 12 & 5.01 & 60 & 155 & 10\\
\rule[-2mm]{0mm}{6mm} 353 & 0.0124 & 43 & 4.86 & 208 & 260 & 17\\
\rule[-2mm]{0mm}{6mm} 545 & 0.0149 & 257 & 4.84 & 1243 &1292 & 86\\
\rule[-2mm]{0mm}{6mm} 857 & 0.0155 & 6828.2 & 4.63 & 31614 & 33039 & 2202\\
\bottomrule
\end{tabular}
\caption{Instrumental sensitivities adopted in this work. The values for {\em Planck} are taken from table 1 of P13 and table 6 of \protect\cite{Planck13-I}.  In the last two columns, we assume hypothetical experiments with the same bandpass filters as {\em Planck} but different, frequency-independent $\sigmapix$ (in $\MJysr$) and $\thetaFWHM$ (in arcmin).  We assume various values for $\sigmapix$ and $\thetaFWHM$ in Section~\ref{sec:fom} and Figure~\ref{fig:fom},  and the corresponding sensitivities can be estimated by rescaling the numbers in the last two columns.}
\label{tab:sensitivity}
\end{table*}

We assume that the detector noise only contributes to the auto power spectra. Based on \cite{Knox95}, the angular power spectrum contributed by the detector noise is given by 
\beq
C_{\ell}^{\rm noise} = w^{-1} {\rm e}^{\ell^2 \sigma_{\rm b}^2} \ ,
\eeq
where $w$ is the weight per solid angle,
\beq
w^{-1} = \sigmapix^2 \Omega_{\rm pix} \ ,
\eeq
 $\Omega_{\rm pix}$ is the solid angle of the pixel, 
\beq
\Omega_{\rm pix} = \thetaFWHM^2 \ ,
\eeq
and $\sigma_{\rm b}$ is the beam size,
\beq
\sigma_{\rm b} = \frac{\thetaFWHM}{\sqrt{8\ln 2}} \ .
\eeq
Here $\sigmapix$ is the detector noise per pixel, and $\thetaFWHM$ is the full width at half-maximum of the Gaussian beam of a given band.  In Table~\ref{tab:sensitivity}, we list the values of $\sigmapix$ and $\thetaFWHM$ of {\em Planck},  as well as the sensitivity $w^{-1/2} = \sigmapix \thetaFWHM$ in the unit of $\uKCMBarcmin$.

In Sections~\ref{sec:fom} and~\ref{sec:comb}, when calculating the constraints for future surveys, we assume that all frequency bands have the same $\thetaFWHM$ and $\sigmapix$ (in the unit of $\MJysr$)\footnote{In the case of {\em Planck}, the 217 GHz band has the lowest detector noise among the CFIRB bands (see Table~\ref{tab:sensitivity}).  When we need to assume a frequency-independent sensitivity, we use the detector noise of 545 GHz (0.015 $\MJysr$) as a conservative baseline.  We also note that the noise levels of 217 and 353 GHz are calibrated in the unit of $\rm\mu K_{CMB}$, while those of 545 and 857 GHz are calibrated in the unit of $\MJysr$. We use the conversion provided by P13, but we note that the conversion depends on the bandpass filter, which will be different for different experiments.}, 
and that the highest multipole is determined by
\beq
\Lmax = \frac{{\rm\pi}}{\thetaFWHM} \ .
\eeq
In the last two columns of Table~\ref{tab:sensitivity}, we list the sensitivities of two hypothetical experiments with the same bandpass filters as {\em Planck} but different $\sigmapix$ and $\thetaFWHM$.  In Section~\ref{sec:fom} and Figure~\ref{fig:fom}, we will assume several combinations of $\sigmapix$ and $\thetaFWHM$, and the corresponding sensitivities can be estimated by rescaling the numbers in the last two columns of Table~\ref{tab:sensitivity}.

\section{Fisher matrix and principal component approach}\label{sec:pca}

\begin{figure*}
\centerline{\includegraphics[width=2\columnwidth]{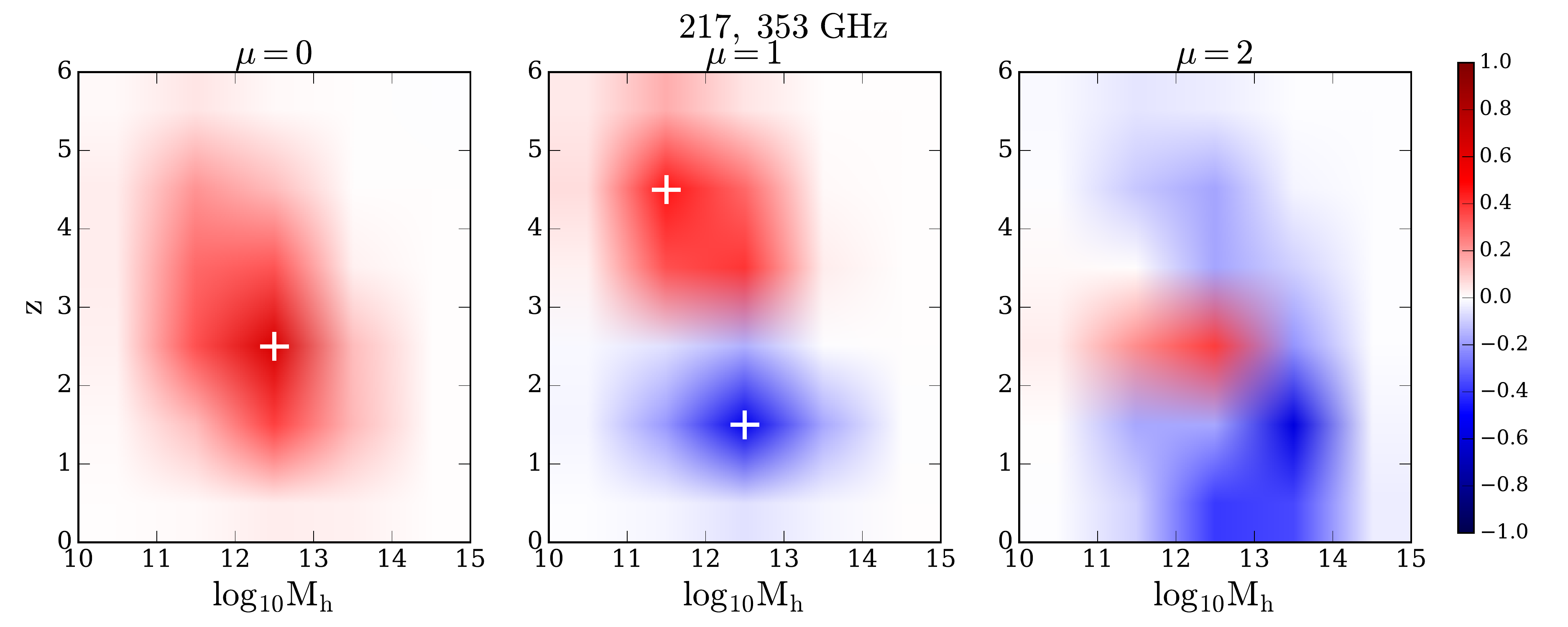}}
\centerline{\includegraphics[width=2\columnwidth]{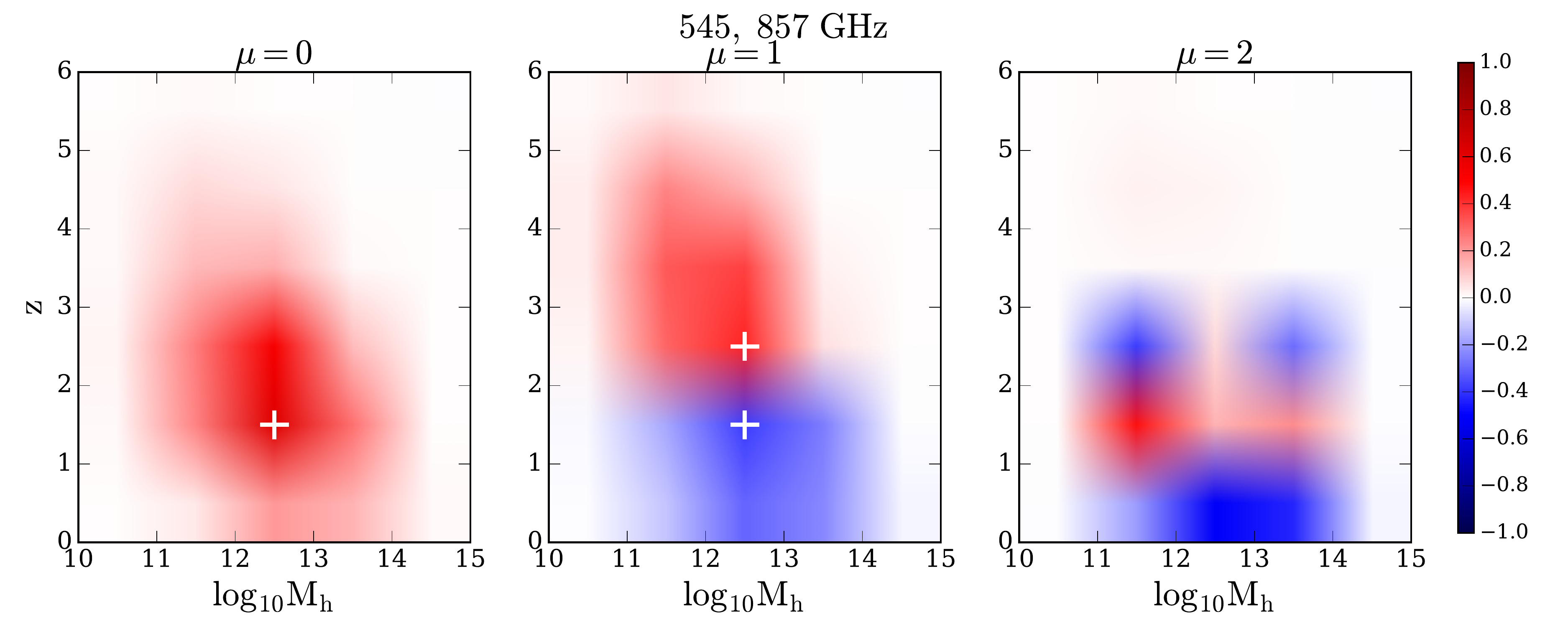}}
\caption{First three principal components of $\SFR(M,z)$ constrained by the CFIRB angular power spectra measured by {\em Planck}. {Top row}: the 217 and 353 GHz bands. The best-constrained redshift lies between $z=2$ and $3$.  {Bottom row}: the 545 and 857 GHz bands. The best-constrained redshift lies between $z=1$ and $2$.  In both cases, the best-constrained halo mass is between $10^{12}$ and $10^{13} \Msun$.}
\label{fig:pca}
\end{figure*}

In order to assess the SFR constraints of future experiments in a model-independent way, we adopt a piecewise parametrization for $\SFR(M,z)$ and calculate the Fisher matrix.  The basic idea is as follows:  for a given range of $M$ and $z$, we use a free parameter to describe the value of SFR in this range.  We then calculate the Fisher matrix and the principal components of these parameters.  The Fisher matrix indicates the information content of the parameters, and the first few principal components indicate the combinations of parameters that are best-constrained.  We note that this approach has been widely applied in cosmology \citep[e.g.,][]{HutererStarkman03,MortonsonHu08}.

\subsection{Piecewise parametrization of SFR}\label{sec:piecewise}

We assume a binned, piecewise parametrization for SFR,
\beqa
\SFR(M,z) =& \sum_i \sum_j\ (1+f_{ij})\ \SFR_{\rm fid}(M,z) \\
&\times H(M_i,M_{i+1}) H(z_j,z_{j+1}) \ ,
\eeqa
where $H(x_i, x_{i+1})$ is the top-hat function and equals 1 if $ x_i \leq x < x_{i+1}$ and 0 otherwise.
The fiducial model $\SFR_{\rm fid}(M,z)$ is described in Section~\ref{sec:Lnu}.  Each of the $f_{ij}$ parameter corresponds to the perturbation around the fiducial model in a mass and redshift bin. We adopt six redshift bins between $z=0$ and $6$, and five logarithmic mass bins between $\log_{10}M$ = 10 and 15; this results in 30 $f_{ij}$ parameters. 

\subsection{Fisher matrix}\label{sec:Fisher}

The Fisher matrix is the inverse of the covariance matrix of model parameters and is used to estimate the parameter constraints for a given data set.  The Fisher matrix for the CFIRB angular power spectra is given by \citep[see, e.g.,][]{Knox01},
\beq
F_{\alpha\beta} = \sum_{\rm binned\ \ell} \frac{(2\ell+1)\Delta\ell f_{\rm sky}}{2} {\rm Tr} \bigg[ 
\bf{C_\ell}^{-1} \frac{{\rm\partial}\bf{C_\ell}}{{\rm\partial}\theta_\alpha}
\bf{C_\ell}^{-1} \frac{{\rm\partial}\bf{C_\ell}}{{\rm\partial}\theta_\beta}\bigg] \ , 
\label{eq:Fisher}
\eeq
where $\bf{C_\ell}$ is a $\Nfreq\times\Nfreq$ matrix ($\Nfreq$ is the number of frequency bands) with its elements given by
\beq
{C_\ell}^{(ij)} = C_{\ell, \nu_i\nu_j} \ ,
\eeq
where
\beq
C_{\ell, \nu_i\nu_j} = C_{\ell, \nu_i\nu_j}^{\rm halo} + C_{\nu_i\nu_j}^{\rm shot} + C_{\ell, \nu_i\nu_j}^{\rm noise}\delta_{ij} \ .
\eeq
The model parameters $\theta_\alpha$ correspond to the binned parameters $f_{ij}$ defined in Section~\ref{sec:piecewise}. We use the same binning of $\ell$ as in P13 (see their table 4).

The fiducial value of each $f_{ij}$ is 0, and we use a broad prior $\sigma_i$ = 1 for each parameter; that is, we add an identity matrix to the Fisher matrix. We then compute the eigenvalues $w_{\mu}$ of the Fisher matrix (sorted from large to small) and the corresponding eigenvectors $S_{\mu}$.  The eigenvector associated with the largest eigenvalue is the first principal component, and it corresponds to the combination of parameters that is best constrained.  The constraint on the $\mu^{\rm th}$ principal component is given by $\sigma_\mu = w_{\mu}^{-1/2}$. We define a principal component to be well-constrained if it has a constraint $\sigma_\mu < 0.1$.

To assess the constraining power of a given survey design, we define the figure of merit (FoM) using the determinant of the Fisher matrix,
\beq
\log_{10} \FoM = \frac{\log_{10}\det({\bf F}_{survey}) - \log_{10}\det({\bf F}_{\rm Planck}) }{30} \ ,
\label{eq:fom}
\eeq
where the denominator 30 corresponds to the number of free parameters in our model.  With this definition, $\log_{10} \FoM = 1$ indicates that the constraints on $\SFR(M,z)$ parameters are on average improved by an order of magnitude.

\subsection{Principal components of SFR from {\em Planck}}

Figure~\ref{fig:pca} presents the first three principal components of the $\SFR(M,z)$ parameters from the CFIRB power spectra measured by {\em Planck}. We arrange the 30 parameters on a two-dimensional grid, and the colour scheme shows the weight on each parameter (white represents 0, and red/blue represents positive/negative values).  The top row corresponds to the principal components from the two low-frequency bands, 217 and 353 GHz, while the bottom row corresponds to the two high-frequency bands, 545 and 858 GHz.  For the former, the peak of the first principal component is at $2<z<3$, while for the latter, the peak of the first component is at $1<z<2$.  This trend confirms that CFIRB is dominated by galaxies between $1<z<3$ \citep[e.g.,][]{Bethermin13}, and that the lower frequency bands probe higher redshifts.  For both cases, the best-constrained halo mass is at $10^{12}\Msun< M<10^{13}\Msun$, which corresponds our assumption of $M_{\rm eff} = 10^{12.6}\Msun$ (see Equation~\ref{eq:SigmaM}). 

We note that the constraints from the high-frequency bands are stronger, and if we combine all four bands, the best-constrained redshift will be at $1<z<2$. For the same reason, if we further add the 3000 GHz band from {\em IRAS} as P13 did, the first principal component will peak at $0<z<1$. Since the 3000 GHz band mainly constrains SFR at $z<1$, we do not include this band in our work.

\begin{figure*}
\includegraphics[width=2\columnwidth]{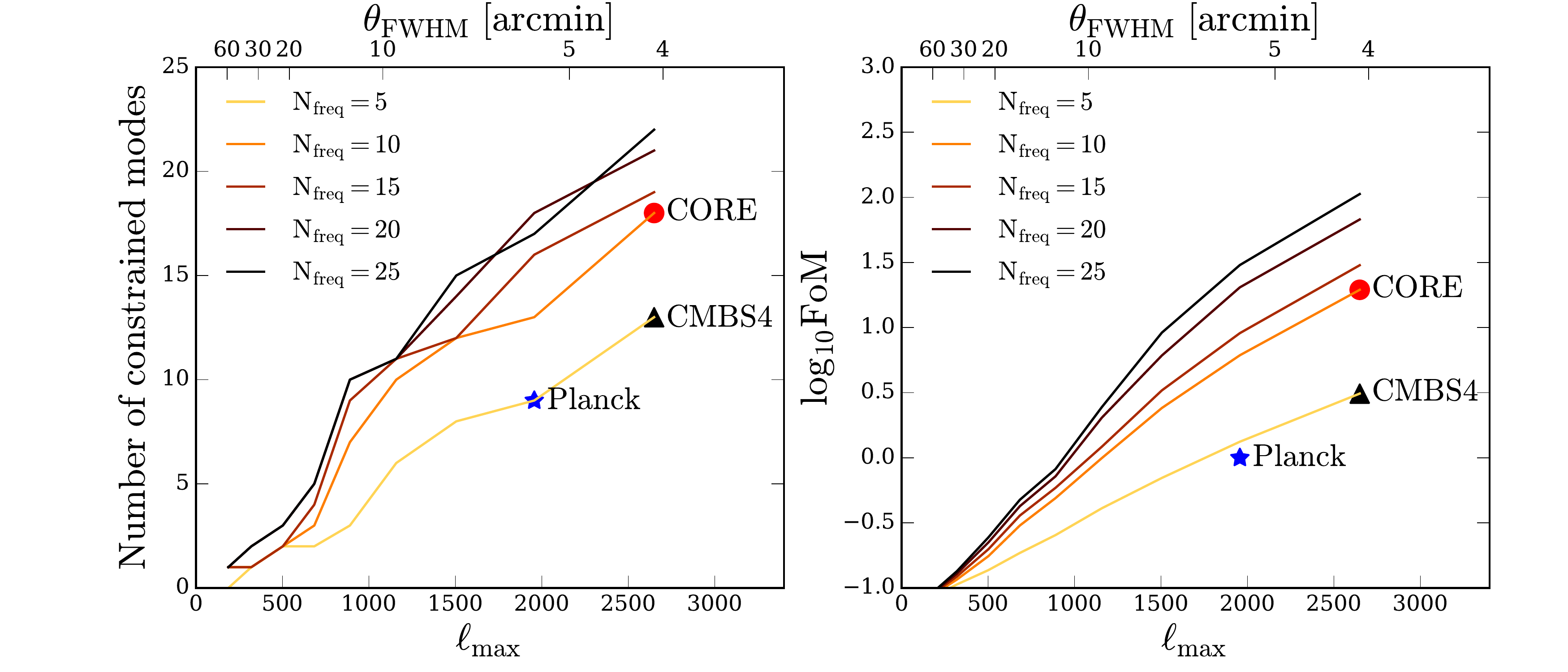}
\includegraphics[width=2\columnwidth]{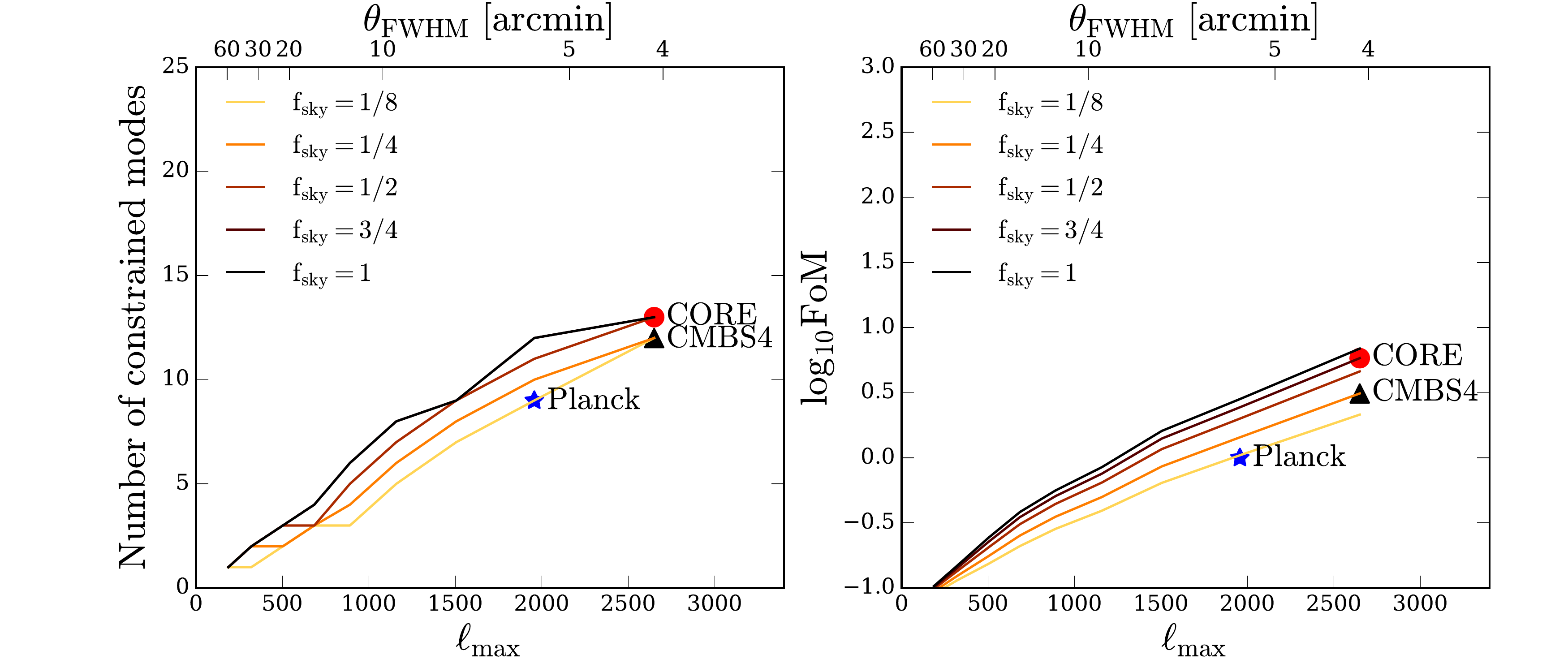}
\includegraphics[width=2\columnwidth]{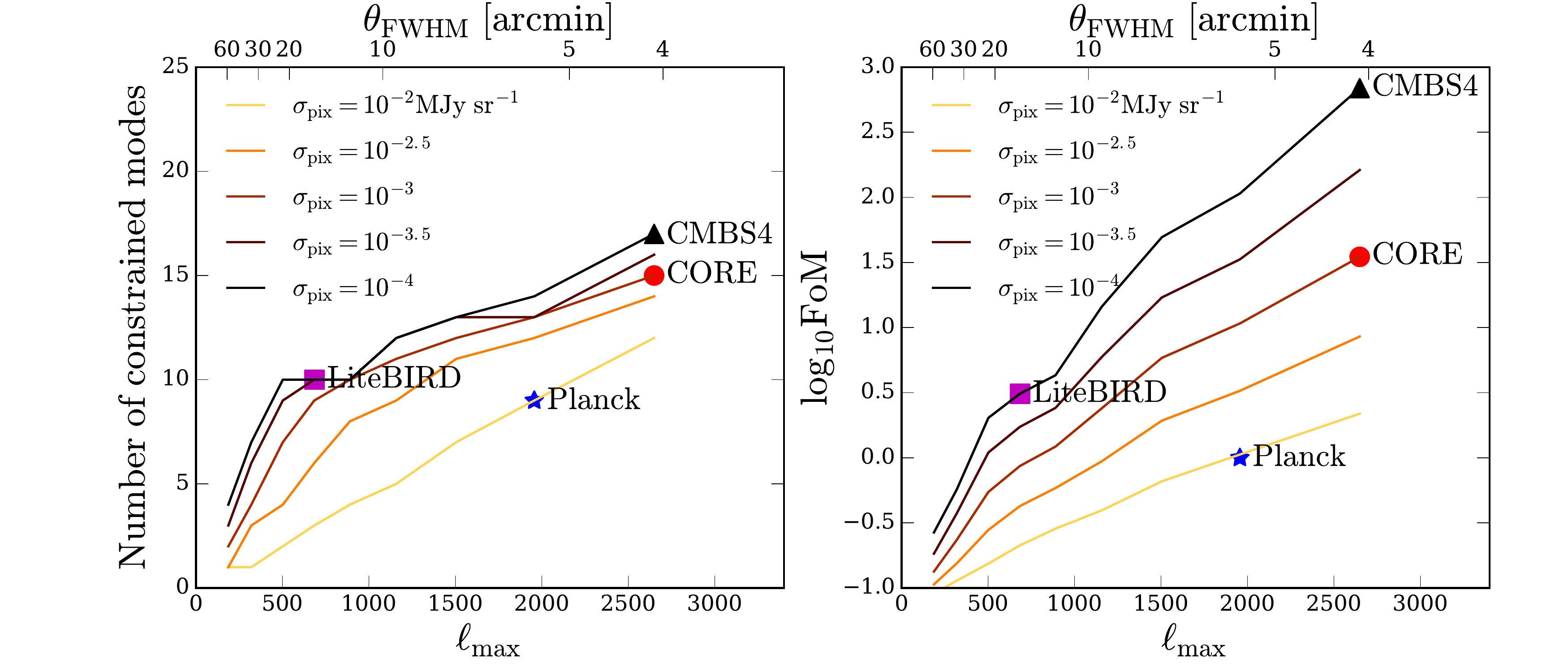}
\vspace{-0.2cm}
\caption{Impact of survey design on the $\SFR(M,z)$ constraints.
{Left}: number of well-constrained modes versus angular resolution.
{Right}: FoM (Equation~\ref{eq:fom}) versus angular resolution.
The baseline survey assumption is presented in Section~\ref{sec:fom}.
{Top}: varying the number of frequency bands logarithmically spaced between 217 and 857 GHz.
{Middle}: varying the sky coverage.
{Bottom}: varying the instrumental sensitivity.
As can be seen, improving the angular resolution and the instrumental sensitivity can significantly improve the $\SFR$ constraints. Increasing the number of frequency bands is only effective when the angular resolution is better than $\approx$ 20 arcmin. In each panel, we mark the approximate loci for {\em Planck}, {\em CORE}, {\em LiteBIRD}, and CMB-S4. We note that the CMB-S4 is likely to have a higher resolution than presented here, and that {\em LiteBIRD} and CMB-S4 will only have 1 or 2 bands ($\lesssim$ 300 GHz) for CFIRB.}
\label{fig:fom}
\end{figure*}
\begin{table*}
\centering
\setlength{\tabcolsep}{0.5em}
\begin{tabular}{cccccc}
\toprule
\rule[-2mm]{0mm}{6mm}  
\multirow{2}[3]{*}{Experiment} & $\thetaFWHM$ & Frequency & Sensitivity & Sensitivity improvement & SFR constraints \\ 
						 & [arcmin] & [GHz] & [$\uKarcmin$] & w.r.t.\ {\em Planck} & $\log_{10}\FoM$ \\ \hline
\rule[-2mm]{0mm}{6mm} {\em CORE} & 1.3--4.7 & 225-795 (9 bands) & 27 (315GHz) & $\sim$ 10$\times$ & 1--1.5 \\
\rule[-2mm]{0mm}{6mm} {\em LiteBIRD} & 16 & 280    & 32   & $\sim$ 100$\times$ & 0.5 \\
\rule[-2mm]{0mm}{6mm} CMB-S4 & 1--3 & 250    & 1     & $\sim$ 100$\times$ & 0.5--3\\
\bottomrule
\end{tabular}
\caption{Designs of future experiments assumed in this work, adopted from \protect\citet[][for {\em CORE}]{CORE11}, \protect\citet[][for {\em LiteBIRD}]{LiteBIRD}, and \protect\citet[][for CMB-S4]{Abazajian16}. The last column shows the improvement of SFR constraints compared with {\em Planck}; $\log_{10}\FoM=1$ corresponds to on average an order of magnitude improvement for SFR for all halo masses and redshifts.}
\label{tab:designs}
\end{table*}

\section{Optimizing the survey design}\label{sec:fom}

We explore how the SFR constraints depend on the survey design. Our baseline survey assumption is as follows:
\begin{itemize}
\item Bands: 217, 353, 545, 857 GHz (1382, 849, 550, 350 $\micron$)
\item Survey area: 2240 $\deg^2$
\item Detector noise: $\sigmapix$ = 0.015 $\MJysr$
\end{itemize}
Figure~\ref{fig:fom} shows how the SFR constraints depend on these factors, and each row corresponds to varying one of these factors. In each panel, we present the SFR constraints ($y$-axis) as a function of angular resolution ($x$-axis, $\Lmax= {\rm\pi}/\thetaFWHM$). The left-hand column corresponds to the number of well-constrained modes (with $\sigma_\mu < 0.1$), while the right-hand column corresponds to the FoM defined in Equation~\ref{eq:fom}. We also mark the approximate loci for several experiments (see Table~\ref{tab:designs} and Section~\ref{sec:discussions}).

\subsection{Number of frequency bands}

The first row of Figure~\ref{fig:fom} corresponds to varying the number of frequency bands ($\Nfreq$) logarithmically spaced between 217 and 857 GHz. As can be seen, when the resolution is better than $\approx$ 20 arcmin ($\Lmax\gtrsim500$), increasing $\Nfreq$ from 5 to 10 leads to approximately an order of magnitude improvement in SFR constraints ($\Delta \log_{10}\FoM = 1$). The improvement with $\Nfreq$ is less significant beyond 10 bands, and the constraints saturate at approximately 20 bands. However, when the angular resolution is worse than $\approx$ 20 arcmin, increasing the number of bands does not significantly improve the SFR constraints.

At a fixed $\Nfreq$, we can see that the SFR constraints sensitively depend on the angular resolution. For example, increasing the angular resolution from $\approx$ 20 arcmin ($\Lmax\approx500$) to $\approx$ 4 arcmin ($\Lmax\approx2700$) can lead to 1.5 orders of magnitude improvement in SFR in the case of $\Nfreq=5$, and 2.5 orders of magnitude in the case of $\Nfreq=25$. Therefore, the angular resolution is more important than the number of bands 
in determining the constraining power of an experiment. The reason is as follows: the SED is a smooth function of frequency and redshift, and thus oversampling in frequency does not significantly increase the information content.  Increasing the angular resolution, on the contrary, increases the number of multipole modes and can significantly improve the SFR constraints \citep[see, e.g.,][for an analogy in the case of galaxy power spectrum]{WuHuterer13}.

The different designs of {\em CORE} \citep{CORE11} and {\em PIXIE} \citep{Kogut11} provide an example of the trade-off between number of bands and angular resolution.  While {\em PIXIE} is designed to densely sample the frequency space (400 bands from 30 to 6000 GHz) with a low angular resolution ($2^{\circ}.6$), {\em CORE} is designed to have fewer bands (15 bands from 45 to 795 GHz) with a higher angular resolution (1.3--4.7 arcmin). Based on our calculation, a survey design like {\em PIXIE} has limited constraining power for SFR, while a survey design like {\em CORE} can significantly improve the SFR constraints beyond {\em Planck}. In fact, {\em PIXIE} is optimized for precise measurement of the absolute intensity and linear polarization of CMB rather than for measurement of SFR, and these two measurements require very different survey strategies.

\subsection{Sky coverage}

The second row of Figure~\ref{fig:fom} corresponds to varying the fraction of the sky coverage, $\fsky$. We show SFR constraints for $\fsky = 1/8, 1/4, 1/2, 3/4$ and $1$, as a function of angular resolution. As can be seen, the improvement due to increased sky coverage is relatively modest compared with increasing angular resolution or $\Nfreq$. For example, increasing the sky coverage from $1/8$ to full sky will lead to approximately half an order of magnitude improvement in SFR constraints.

We note that the {\em Planck} 2013 CFIRB results are based on a survey area of 2240 deg$^2$ ($\fsky \approx 0.05$). This sky coverage is limited by the radio measurements for the neutral atomic hydrogen ({\sc Hi}) column density, which are required to remove the foreground Galactic dust. Increasing the sky coverage will require new radio surveys and thus substantial extra observational resources \citep[e.g.,][]{HI4PI16}; in this sense, increasing the sky coverage may not be the most efficient way for improving the SFR constraints from CFIRB. Nevertheless, increasing the sky coverage and observing different regions of the sky is still invaluable because it provides essential consistency checks and controls systematic errors.

\subsection{Instrumental sensitivity}

The third row of Figure~\ref{fig:fom} corresponds to different instrumental sensitivities. We characterize the sensitivity using the detector noise ($\sigmapix$) and the angular resolution ($\thetaFWHM$) introduced in Section~\ref{sec:detector}. For $\sigmapix$, we use the intensity unit $\MJysr$; in Table~\ref{tab:sensitivity} we present the conversion between $\MJysr$ and $\uKCMB$, as well as the sensitivity in the commonly used unit $\uKCMBarcmin$ for the {\em Planck} bands. In the last two columns of Table~\ref{tab:sensitivity}, we list two hypothetical experiments with given $\sigmapix$ and $\thetaFWHM$, and we calculate the corresponding sensitivity in the unit of $\uKCMBarcmin$ assuming the bandpass filters of {\em Planck}. For other combinations of $\sigmapix$ and $\thetaFWHM$, the sensitivity can be estimated by rescaling the numbers in the last two columns. 

We assume that all frequency bands have the same sensitivity, and we show the cases of $\sigmapix = 10^{-2}, 10^{-2.5}, 10^{-3}, 10^{-3.5},$ and $10^{-4}\ \MJysr$. For comparison, {\em Planck} has $\sigmapix = 0.0155\ \MJysr$ at 857 GHz. As can be seen, reducing the pixel noise can significantly improve the SFR constraints; for example, reducing the pixel noise by two orders of magnitude can improve the SFR constraints by 1--2.5 orders of magnitude, and the improvement is greater with a higher angular resolution. 

We can only roughly estimate the pixel noise levels for future experiments. In Table~\ref{tab:designs}, we list the targeted sensitivities for several experiments, taken from \citet[][for {\em CORE}]{CORE11}, \citet[][for {\em LiteBIRD}]{LiteBIRD} and \citet[][for CMB-S4]{Abazajian16}. From these values, we estimate that {\em CORE} will have a sensitivity $\sim$ 10 time better than {\em Planck}, and that {\em LiteBIRD} andCMB-S4 will have sensitivities $\sim$ 100 times better than {\em Planck}.  In Figure~\ref{fig:fom}, we mark the approximate loci of these experiments accordingly. The CMB-S4 is more likely to have even higher angular resolution and sensitivity than presented in our figure.

We note that CMB-S4 and {\em LiteBIRD} will include only the low-frequency bands for CFIRB ($\lesssim$ 300 GHz). Although these bands have limited information for SFR at $z\lesssim2$, they will provide constraints on high-redshift galaxies ($z\gtrsim3$). Such constraints will be invaluable, because at $z\gtrsim3$ CFIRB may never be completely resolved into individual galaxies. However, at these low-frequency bands, the power spectrum is dominated by CMB, and separating CMB and CFIRB can be a major challenge (see P13). 

\section{Optimal range of frequency bands}\label{sec:comb}

\begin{figure*}
\centerline{\includegraphics[width=2.2\columnwidth]{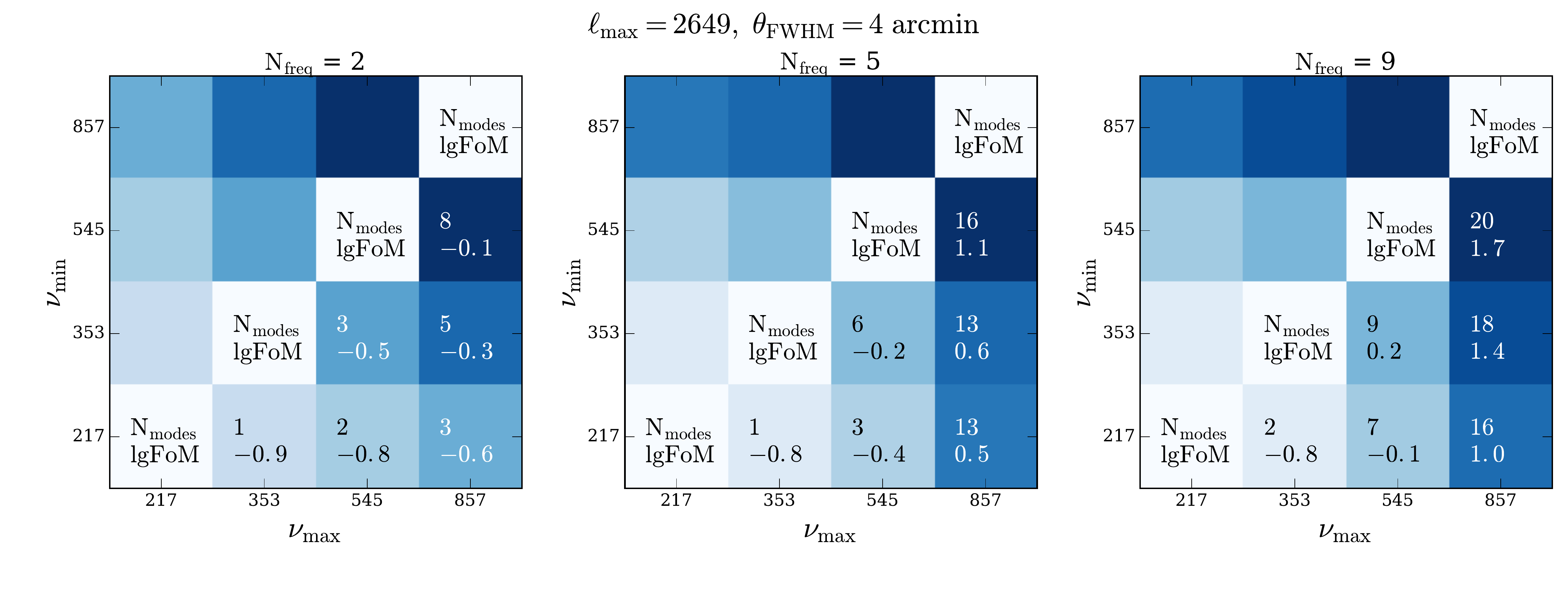}}
\centerline{\includegraphics[width=2.2\columnwidth]{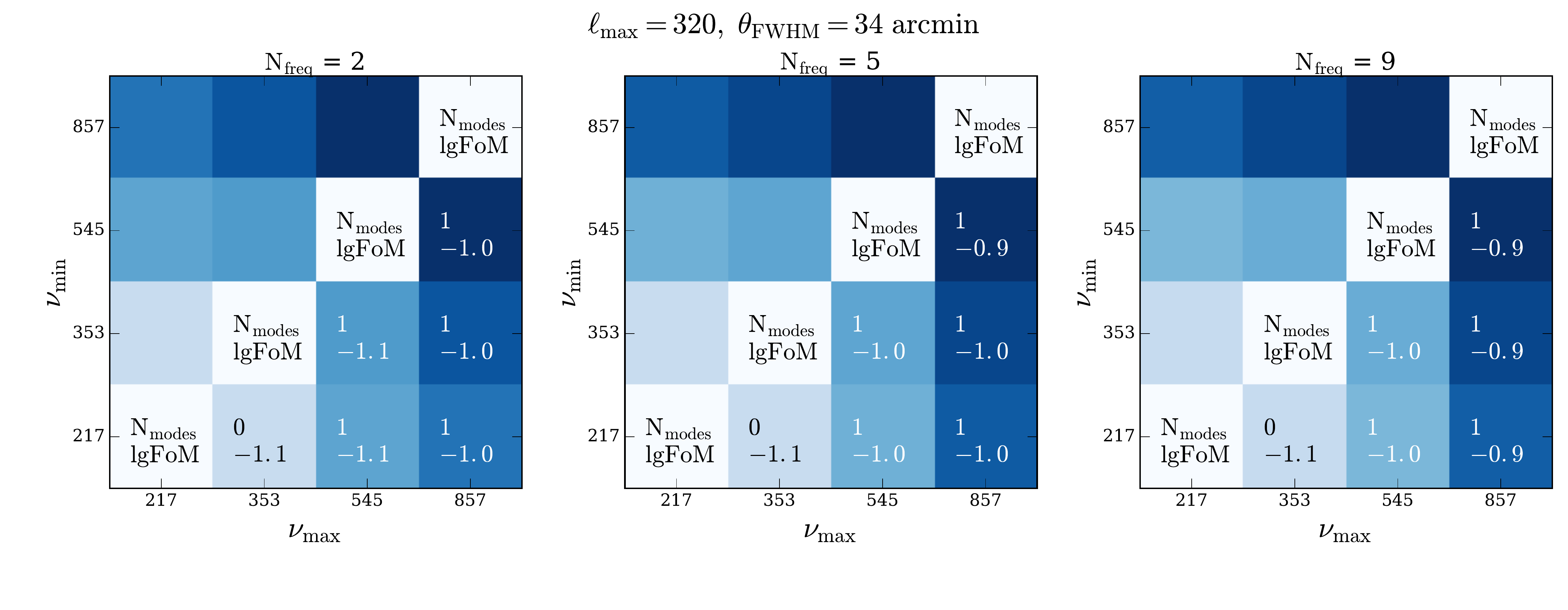}}
\caption{Comparison of various ranges and numbers of frequency bands.  The top/bottom row corresponds to high/low angular resolution.  The left-hand/central/right-hand column corresponds to 2/5/9 bands logarithmically spaced between $\numin$ ($y$-axis) and $\numax$ ($x$-axis).  In each cell, the shading corresponds to the $\log_{10}\FoM$, and we show the number of well-constrained modes and the $\log_{10}\FoM$.  The high-frequency bands can constrain more modes, but these modes are limited to low redshift. When the angular resolution is high, increasing the number of bands is effective in improving the SFR constraints, and the best improvement is when $\numin$ and $\numax$ span a wide range.  However, when the angular resolution is low, increasing the number of bands cannot significantly improve SFR constraints.}
\label{fig:comb}
\end{figure*}

In this section, we explore the SFR constraints from various ranges of frequency bands. We use ($\numin$, $\numax$) pairs selected from the {\em Planck} bands: (217, 353, 545, 857) GHz, and we assume $N_{\rm freq}$ logarithmically spaced bands between $\numin$ and $\numax$. We adopt the same baseline survey design as in Section~\ref{sec:fom}.

Figure~\ref{fig:comb} shows the comparison between different $N_{\rm freq}$ and ($\numin$, $\numax$). The top row corresponds to a high-resolution survey with $\thetaFWHM = 4$ arcmin ($\Lmax=2649$), while the bottom row corresponds to a low-resolution survey with $\thetaFWHM = 34$ arcmin ($\Lmax=320$). The left-hand, central, and right-hand panels correspond to $N_{\rm freq} = $ 2, 5, and 9. For each panel, the $y$-axis corresponds to $\numin$, and the $x$-axis corresponds to $\numax$. In each cell, the two numbers correspond to the number of well-constrained modes and the $\log_{10}\FoM$ from the given frequency bands, and the shading is based on $\log_{10}\FoM$. 

From the top row, we can see that for a given number of bands, the high-frequency bands always constrain the largest number of modes. However, as shown in Figure~\ref{fig:pca}, the high-frequency bands only constrain SFR at low redshift. In order to constrain SFR at high redshift, we need low-frequency bands, but the improvement is rather slow when we increase the number of bands between 217 and 353 GHz. On the other hand, if we compare the improvements associated with increased $\Nfreq$, we can see that the cells correspond to (217, 857) and (353, 857) show the most significant improvement; that is, spanning a wide range of frequencies can be beneficial.

From the bottom row, we can see that the constraining power of an experiment is significantly reduced when the resolution is low.  The FoM values are low with this resolution, and increasing the number of bands barely improves the constraining power of the experiment. Comparing the top and bottom panels, we can once again see that it is more important to improve the angular resolution than to increase the number of bands for constraining SFR.

\section{Implications for future CFIRB experiments}\label{sec:discussions}

In this section, we put our results in the context of several future experiments that are currently being planned.  We note that most of these experiments are optimized for measuring CMB polarization, and they include FIR/submm bands for measuring dust emissions in order to control systematics.  
These bands will contain rich information for extragalactic astrophysics, and the CFIRB measured from these experiments will significantly improve the constraints on the cosmic star-formation history.  Below we compare the survey designs of several experiments (also see Table~\ref{tab:designs}), and we note that the design we quote below are only approximate and are still evolving.

The {\em CORE} mission is the next generation space-borne CMB experiment, which is proposed to the European Space Agency and designed to be the successor of {\em Planck}.  The major improvement beyond {\em Planck} is the high sensitivity and a diffraction-limited resolution, as well as twice as many frequency bands as {\em Planck}.  One of the experimental designs, as presented in  \cite{CORE11}, includes 15 frequency bands between 45 GHz (6.7 mm) and 795 GHz (377 $\micron$), and the angular resolution ranges from 23 arcmin at 45 GHz to 1.3 arcmin at 795 GHz.  In addition, the sensitivity of {\em CORE} is expected to be 10--30 times better than that of {\em Planck}.  As we have shown in Section~\ref{sec:fom}, such a survey design is very close to optimal for constraining SFR. 

The {\em CORE} team has recently published several designs, including a mirror ranging from 1 to 1.5 m, and $\numax$ from 600 to 800 GHz \citep[][]{CORE15,CORE16a,CORE16b}.  Based on our results, if one has to choose between the number of bands and the angular resolution, it is the latter that will provide stronger SFR constraints.  We note that the extragalactic sources that will be resolved by {\em CORE} have been studied in  \cite{CORE15,CORE16a}. Our work is highly complementary to those studies in the sense that we focus the SFR constraints of faint, unresolved galaxies.

The {\em LiteBIRD} mission is another next-generation space-borne CMB experiment, which is designed to cover the frequency range from 50 to 320 GHz, with a sensitivity of 2 $\uKarcmin$ and an angular resolution of 16 arcmin \citep{LiteBIRD}.  The sensitivity is superior to CORE, while the angular resolution is lower.  As we have shown in Section~\ref{sec:fom}, such a high sensitivity is beneficial for improving the SFR constraints from CFIRB.  The highest frequency is relatively low for CFIRB, but it can be beneficial for constraining SFR at high redshift.

The next-generation ground-based CMB-S4 experiment will have superior angular resolution ($\lesssim$ 1 arcmin) and a sensitivity two orders of magnitude better than {\em Planck} \citep{Abazajian16}.  Operating from the ground, CMB-S4 can only observe through the atmospheric windows, and the highest frequency is around 250 GHz.  Therefore, CMB-S4 will provide CFIRB measurements at low frequencies and constrain the SFR at $z\gtrsim3$.  In addition, CMB-S4 will measure CMB lensing to unprecedented precision,  enabling detailed studies of the cross-correlation between CFIRB and CMB lensing potential. This correlation is essential for probing the connection between SFR and halo mass and providing consistency checks for the CFIRB power spectra.  Nevertheless, at $\lesssim$ 250 GHz, separating CFIRB from CMB can be a major challenge, and careful component separation will be required to extract CFIRB to high precision. 

The {\em PIXIE} (\citealt{Kogut11}) is a space-borne CMB mission designed for measuring the large-scale polarization.  It will significantly improve the precision of the measurement of the absolutely intensity of the cosmic background radiation compare with {\em COBE}.  The survey design includes 400 effective channels from 30 GHz (10 mm) to 6 THz (50 $\micron$), with an angular resolution of $2^{\circ}.6$. Despite its unique role in measuring the absolute intensity of CFIRB, it has limited angular resolution to measure the anisotropies of CFIRB; therefore, as we have demonstrated, it is not optimal for constraining SFR.

In addition, FIR/submm survey telescopes with wide field of view have also been planned, including the Cerro Chajnantor Atacama Telescope (CCAT; \citealt{CCAT}) and the Far-infrared Surveyor (Origins Space Telescope; \citealt{Meixner16}). These telescopes usually operate at frequencies higher than CMB telescopes ($\gtrsim$ 500 GHz) and will also measure CFIRB in much smaller area of the sky.  More importantly, these telescopes will have the power to resolve a significant fraction of the sources contributing to CFIRB at low redshift, and thus they are complementary to the CFIRB experiments discussed in this work.

\section{Summary}\label{sec:summary}

We explore the optimal survey strategies for future CFIRB experiments, and our goal is to maximize the SFR constraints extracted from the CFIRB angular power spectra. We introduce a model-independent, piecewise parametrization for $\SFR(M,z)$, and we calculate the Fisher matrix to estimate how well future experiments can constrain SFR. Our findings are summarized as follows.

\begin{itemize}

\item From the first principal component of the Fisher matrix, we find that for the CFIRB power spectra observed by {\em Planck},  the 217 and 353 GHz bands (1382 and 849 $\micron$) provide the strongest SFR constraints at $2<z<3$, while the 545 and 857 GHz bands (550 and 350 $\micron$) provide the strongest SFR constraints at $1<z<2$.  In all redshifts, the strongest SFR constraints are associated with halo mass between $M=10^{12}$ and $10^{13} \Msun$.

\item We find that the angular resolution and the instrumental sensitivity are the most important factors for improving SFR constraints.  For example, improving the angular resolution from 20 to 4 arcmin can lead to 1.5 to 2.5 orders of magnitude improvement in SFR constraints.  In addition, improving the sensitivity by 10 or 100 times with respect to {\em Planck} can lead to one to two orders of magnitude improvement in SFR constraints, and the improvement is greater when the angular resolution is higher.  We also find that increasing the survey area only leads to modest improvement in SFR constraints.

\item Increasing the number of frequency bands is not always effective in improving SFR constraints.   With an angular resolution similar to {\em Planck} ($\approx$ 5 arcmin), doubling the number of frequency bands of {\em Planck}  can improve SFR constraints by an order of magnitude. Further increasing the number of bands has less impact on SFR constraints, and the constraints saturate at approximately 20 bands.   However, if the angular resolution is lower than 20 arcmin, increasing the number of bands can hardly improve the SFR constraints.   Therefore, in terms of maximizing the SFR constraints, a survey design with  high angular resolution and relatively fewer bands (e.g., {\em CORE}) is favoured over  a design with low angular resolution and a large number of bands (e.g., {\em PIXIE}).

\item We explore the constraining power of various ranges and numbers of frequency bands. With an angular resolution similar to {\em Planck}, we find that the SFR constraints improve relatively slowly when we increase the number of low-frequency bands; that is, it is relatively difficult to improve SFR constraints at high redshift.  If we consider SFR constraints for all redshifts, spanning a wide range of frequencies will lead to the most significant improvement.   On the other hand, when the resolution is low ($\lesssim$ 20 arcmin), increasing the number of bands in any frequency range has very limited impact on SFR constraints.

\item When comparing our results with the designs of several future CMB missions, we find that {\em CORE} has nearly the optimal angular resolution and number of bands to significantly improve SFR constraints. In addition, {\em LiteBIRD} will have superior sensitivity, and CMB-S4 will have both superior sensitivity and high angular resolution.  Since {\em LiteBIRD} and CMB-S4 will only cover low frequencies ($\lesssim$ 300 GHz), they will mainly constrain the SFR at high redshift.  The proposed Far-Infrared Surveyor will have superior angular resolution and sensitivity, but it will focus on high frequencies and smaller survey area; thus, it will efficiently resolve sources contributing to CFIRB and constrain the SFR at low redshift.   The design of {\em PIXIE} is optimized for large-scale polarization measurements and is not optimal for constraining SFR. 

\end{itemize}

In this work, we only consider the constraints of SFR from CFIRB angular power spectra.   We expect that other FIR/submm observations will provide extra SFR constraints and important consistency checks. These observations include the IR luminosity functions \citep[e.g.,][]{Gruppioni13}, submm number counts \citep[e.g.,][]{Bethermin13,Valiante16}, the cross-correlation between CFIRB and CMB lensing potential \citep[e.g.,][]{Planck13XVIII}, and the cross-correlation between CFIRB and other extragalactic background light \citep[e.g.,][]{Cooray16}.  Among these observations, CFIRB is still a powerful and unique probe because it provides the rare opportunity to constrain the star formation activities under the current resolution limit.   A robust measurement of CFIRB will require not only superior angular resolution and sensitivity of instruments but also careful control of systematic errors and comprehensive cross-correlation studies.

\section*{Acknowledgements}
HW acknowledges the support by the US National Science Foundation (NSF) grant AST1313037.  The calculations in this work were performed on the Caltech computer cluster Zwicky, which is supported by NSF MRI-R2 award number PHY-096029.  Part of the research described in this paper was carried out at the Jet Propulsion Laboratory, California Institute of Technology, under a contract with the National Aeronautics and Space Administration.

\bibliographystyle{mnras}
\bibliography{/Users/hao-yiwu/Dropbox/master_refs}

\appendix

\section{Derivation of shot noise}\label{app:shotnoise}

The spectral flux is given by
\beq
S_\nu = \frac{L_{(1+z)\nu}}{4{\rm\pi} \chi^2 (1+z)} = S(1+z)\Theta_{(1+z)\nu} \ , 
\eeq
where $S$ is the bolometric flux given by 
\beq
S = \frac{L }{4{\rm\pi}\chi^2 (1+z)^2} \ .
\eeq
The shot noise of the {\em auto} power spectrum is given by the integration
\beqa
C^{\rm shot}_{\nu\nu}
&= \int \dV \int^{S_\nu^{\rm max}}_{S_\nu^{\rm min}} \dS_\nu \frac{\dn}{\dS_\nu} S_\nu^2 \\
&= \int \dV \int^{S^{\rm max}}_{S^{\rm min}} \dS \frac{\dn}{\dS} S^2 (1+z)^2 \Theta^2_{(1+z)\nu} \ , \\
\eeqa
where
\beq
\dV = \frac{D_H}{E(z)}\chi^2 \dz {\rm d}\Omega \ .
\eeq
The upper limit of the integration, $S_\nu^{\rm max}$, corresponds to the flux cut of the observation, and we adopt the values in table 1 of P13.  In practice, our model includes negligible number of galaxies above the flux cut because we do not include starburst or lensed galaxies. 

The shot noise of the {\em cross} power spectrum is given by
\beq
C^{\rm shot}_{\nu\nu'} = \int \dV \int^{S^{\rm max}}_{S^{\rm min}} \dS \frac{\dn}{\dS} S^2 (1+z)^2 \Theta_{(1+z)\nu} \Theta_{(1+z)\nu'} \ .
\eeq

To calculate $\dn/\dS(z)$, we integrate over the probability distribution function of $S$ at a given $M$, 
\beq
\frac{\dn}{\rmd\ln S} = \int d\ln M\ \frac{\dn}{\rmd\ln M} \ P(\ln S | \ln M) \ ,
\eeq
where we assume $P(\ln S | \ln M)$ follows a normal distribution
\beq
P\big(\ln S | \ln M\big) \sim {\rm Normal}\bigg(\avg{\ln S(M)}; \sigma^2\bigg) \ ,
\eeq
with the mean given by
\beq
\avg{\ln S(M)} = \ln \avg{S(M)} - \frac{\sigma^2}{2} \ ,
\eeq
where
\beq
\avg{S(M,z)} = \frac{\avg{\LIR(M,z)}}{4{\rm\pi} \chi^2 (1+z)} \ .
\eeq
We find that a scatter of $\sigma=0.9$ (0.28 dex) is required to reproduce the shot noise in P13.  We note that this scatter does not affect any of the equations presented in Section~\ref{sec:CL}, because those questions only involve $\avg{\LIR}$.  We also find that satellite galaxies have a negligible contribution to the shot noise.

\bsp	
\label{lastpage}
\end{document}